\documentclass{pasj01}
\Received{$\langle$reception date$\rangle$}
\Accepted{$\langle$acception date$\rangle$}
\Published{$\langle$publication date$\rangle$}

\usepackage{xspace}
\usepackage{natbib}
\usepackage{ulem}

\begin{document}

\title{Star Formation Induced by Cloud-Cloud Collisions and Galactic Giant Molecular Cloud Evolution}
\author{Masato I.N. KOBAYASHI\altaffilmark{1},
        Hiroshi KOBAYASHI\altaffilmark{1},
        Shu-ichiro INUTSUKA\altaffilmark{1},
        Yasuo FUKUI\altaffilmark{1}
        }%
\altaffiltext{1}{Division of Particle and Astrophysical Science, Graduate School of Science, Nagoya University, Aichi 464-8602, Japan}
\email{masato.kobayashi@nagoya-u.jp}


\newcommand{\msun}{\mbox{$M_{\odot}$}\xspace}
\newcommand{\eg}{{\it e.g.}\xspace}
\newcommand{\cf}{{\it c.f.}\xspace}
\newcommand{\ie}{{\it i.e.}\xspace}
\newcommand{\mytf}{T_{\rm f}}
\newcommand{\mytsg}{T_{\rm sg}}
\newcommand{\mytd}{T_{\rm d}}
\newcommand{\mytdccc}{T_{\rm d,col}}
\newcommand{\mytsb}{T_{*}}
\newcommand{\mytsd}{T_{\rm dest}}
\newcommand{\mytffid}{T_{\rm f,fid}}
\newcommand{\mytdfid}{T_{\rm d,fid}}
\newcommand{\mytcol}{T_{\rm col}}
\newcommand{\mytcolnum}{T_{\rm col,num}}
\newcommand{\mytcolmass}{T_{\rm col,mass}}
\newcommand{\mytgrow}{T_{\rm growth}}
\newcommand{\mytsteady}{T_{\rm steady}}
\newcommand{\myratio}{\mytf/\mytd}
\newcommand{\mtrunc}{m_{\rm trunc}}
\newcommand{\myrec}{\varepsilon_{\rm res}}
\newcommand{\myrecsteady}{\varepsilon_{\rm res,std}}
\newcommand{\mysfe}{\varepsilon_{\rm SFE}^{}}
\newcommand{\myfred}{f_{\rm red}}
\newcommand{\mcrit}{m_{\rm crit}}
\newcommand{\myftot}{F_{\rm total}}
\newcommand{\myfin}{F_{\rm in}}
\newcommand{\myfout}{F_{\rm out}}
\newcommand{\mydotmg}{\dot{M}_{\rm grow}}
\newcommand{\mydotmd}{\dot{M}_{\rm disp}}
\newcommand{\mymmin}{m_{\rm min}}
\newcommand{\mymmax}{m_{\rm max}}
\newcommand{\icarus}{Icarus}

\maketitle

\begin{abstract}
\label{sec:abst}
Recent radio observations towards nearby galaxies started to map
the whole disk and to identify 
giant molecular clouds (GMCs) even in the regions between
galactic spiral structures. 
Observed variations of GMC mass functions in different galactic environment 
indicates that 
massive GMCs preferentially reside along galactic spiral structures whereas
inter-arm regions have many small GMCs.
Based on the phase transition dynamics from magnetized warm neutral medium to 
molecular clouds, Kobayashi et al. 2017 proposes a semi-analytical evolutionary description 
for GMC mass functions including cloud-cloud collision (CCC) process.
Their results show that CCC is less dominant in shaping the mass function of GMCs 
compared with the accretion of dense HI gas driven by the propagation of supersonic shock waves.
However, their formulation does not take into account 
the possible enhancement of star formation by CCC\@.
Radio observations within the Milky Way 
indicate the 
importance of CCC
for the formation of star clusters and massive stars.
In this article, we reformulate the time evolution equation
largely modified 
from Kobayashi et al. 2017 so that we 
additionally compute star formation subsequently taking place in CCC clouds.
Our results suggest that,
although CCC events between smaller clouds 
are more frequent than the ones between massive GMCs,
CCC-driven star formation is mostly driven by massive GMCs $\gtsim 10^{5.5} \msun$
(where $\msun$ is the solar mass).
The resultant cumulative CCC-driven star formation may 
amount to
a few 10 per cent of the total star formation in the Milky Way and nearby galaxies.

\end{abstract}

\section{INTRODUCTION}
\label{sec:intro}
Giant molecular clouds (GMCs) are believed to be the parental structure of hydrogen molecules (H$_2$)
forming stars, which have a typical mass $\gtsim 10^4 \msun$,
where $\msun$ is the solar mass, 
and $\gtsim 10$ parsec (pc) (\eg, \citealt[][]{Williams2000,Kennicutt2012}).
Understanding of star formation and subsequent galaxy evolution 
therefore requires a framework to describe GMC formation, evolution, and dispersal on galactic scales.
Over the last 10 years, radio observations towards nearby galaxies have started to map GMC distributions 
throughout their entire disks (\citealt[][]{Engargiola2003, Rosolowsky2003, Rosolowsky2007, 
Koda2009, Koda2011, Koda2012, Colombo2014a, Colombo2014b}).
Especially, the Plateau de Bure Interferometer (PdBI) Arcsecond Whirlpool Survey (PAWS) program
observed Galaxy M51 in detail with PdBI and the IRAM 30m telescope \citep{Schinnerer2013}.
One of the highlighted results is the observed variation in GMC mass function (GMCMF) \citep{Colombo2014a};
when fitted with a power-law profile $n_{\rm cl}(m) \propto m^{-\alpha}$
, where $n_{\rm cl}(m)$ represents the differential number density of GMC with mass $m$
(\ie, the cumulative number density 
of GMCs with mass greater than $m$, $n(>m)$, is given as $n(>m) = \int_{m}^{\infty} n_{\rm cl}(m) {\rm d}m$),
the GMCMF shows a shallower slope $-\alpha \sim -1.3$ in arm regions 
whereas $-\alpha \sim -2.6$ in inter-arm regions.
This result indicates that massive GMCs preferentially reside along galactic spiral arms
whereas the mass budget in inter-arm regions is dominated by less massive GMCs ($m<10^{5.5} \msun$).
It is therefore required to connect 
such observed trends in ensemble GMC populations
with the phase transition dynamics on pc and sub-pc scales (as mentioned in the following paragraph)
for understanding GMC evolution as well as subsequent star formation and galaxy evolution.

The interstellar medium (ISM) constructs thermally bistable phases of atomic hydrogen
due to the balance between radiative cooling and photoelectric heating 
(partially cosmic ray heating as well)
(\citealt{Field1969, Wolfire1995, Wolfire2003}).
One of the two phases is 
warm neutral medium (WNM) with the temperature $\sim 6000$ K and the density $\sim$ 0.1 atomic hydrogen cm$^{-3}$, 
and the other phase is 
cold neutral medium (CNM) with the temperature $\sim 100$ K and the density $\sim$ 100 atomic hydrogen cm$^{-3}$,
which is a precursor of GMCs.
WNM occupies most of the volume in galactic disks,
and thus the phase transition dynamics from WNM to CNM is important for GMC formation on galactic scales.
Over the last 20 years, multiphase ISM simulations 
investigate the propagation of shock waves in WNM
to override the pressure balance between WNM and CNM,
and they successfully form molecular clouds from WNM through thermal instability
\citep[\eg][]{Walder1998a,Walder1998b,Koyama2002,Audit2005,Audit2008,Heitsch2005,Heitsch2006b,vazquezsemadeni2006,Hennebelle2007a,Hennebelle2007b}.
Here, 
shock waves are supposed to be driven by 
expanding supernovae or H{\sc ii} regions in the real ISM\@.
However, multiphase magnetohydrodynamics ISM simulations since about 10 years ago
have revealed that magnetic field pressure can support the ISM against compression
driven by shock waves,
so that magnetic fields retard the cloud formation 
from magnetized WNM with a typical field strength of just a few micro Gauss in the ISM
\citep[\eg][]{Inoue2008}.
These results suggest that successful molecular cloud formation takes place 
only after a few 10 multiple shocks 
compress WNM with the shock propagation directions
misaligned with local magnetic field lines.
Therefore, multiple episodes of supersonic compression 
is presumably essential to form molecular clouds from magnetized WNM, and 
the typical phase transition timescale by such multiple compression 
is estimated about a few 10 Myr \citep{Inutsuka2015}.

To connect individual GMC formation governed by multiple episodes of compression 
with the evolution of GMC populations over galactic disks,
\citet{Inutsuka2015} propose a bubble scenario where 
the network of expanding shells
due to expanding supernovae and H{\sc ii} regions
create repeated
supersonic shock propagations.
Based on this paradigm, they formulate a time evolution equation of GMCMF 
due to the multiple episodes of compression and GMC self-dispersal.
However, this formulation 
neglected the change of cloud mass function by cloud-cloud collisions (CCCs).
From 1970s, GMC evolution due to CCCs alone
is extensively investigated 
by coagulation equation 
\citep[\eg,][]{Kwan1979,Scoville1979,Cowie1980}
and N-body simulations \citep[\eg,][]{Levinson1981,Kwan1983,Tomisaka1984,Tomisaka1986}.
Thus, in the previous studies of the cloud mass function, 
the detailed formation/dispersal of clouds and CCC
were studied separately.
These studies are separately conducted but have to be incorporated to 
reveal which process plays a dominant role in shaping what part of GMCMF evolution.

\citet{Kobayashi2017a} extended the formulation in \citet{Inutsuka2015} by including CCCs. 
Their results indicate that CCC modifies only the massive end of GMCMF 
while
GMCMF exhibits a power-law slope in the low mass regime ($\lesssim 10^{5.5} \msun$),
which is well characterized by a combination of two timescales:
formation/growth and dispersal (see section~\ref{sec:previous_formulation} in this article).
CCC determines the power-law slope of the cloud mass function 
only when 
these formation and dispersal are slower processes than CCC
(\eg, such as galactic centers where GMC number densities are higher than disk regions;
\cf, \citealt{Kwan1979,Cowie1980,Tomisaka1984}).
However, \citet{Kobayashi2017a} focus only on gas phase, namely GMCMF, 
so that they have not investigated resultant star formation 
out of those GMC populations. 
Indeed, recent radio observations have found
increasing number of star cluster forming sites likely triggered by CCC
(\eg, \citealt[][]{Torii2011,Nakamura2012,Fukui2014,Torii2015,Fukui2016,
Torii2017a,Torii2017c,
Nishimura2017a,Nishimura2017b,Fukui2017b,Fukui2017c,Fukui2017d,Fukui2017e,
Sano2017a,Ohama2017a,Ohama2017b,Kohno2017,Hayashi2017,Saigo2017,Tsutsumi2017};
see also \citealt{Furukawa2009,Ohama2010,Dobashi2014,Nakamura2014,Fukui2015b,Tsuboi2015,Dewangan2016,Dewangan2017,Ohama2017c,Sano2017b,Torii2017b}).
Their interpretation of CCC
indicates the possible importance of CCC-driven star formation
across the Milky Way galaxy from the solar circle to the Galactic Center.
Simulations of colliding GMCs \citep[\eg,][]{Inoue2013,Takahira2014,Inoue2017,Wu2017c,Takahira2018} also 
suggest molecular cloud core formation in a shocked compressed layer,
which may result in rapid star cluster formation and efficient massive star formation.

In this article, we introduce star formation rate (SFR) implementation to our time-evolution equation for GMCMF
from \citet{Kobayashi2017a},
and calculate 
SFR with a given GMCMF
to evaluate the relative contribution by CCC to total star formation
across galactic disks.
We also evaluate CCC timescales as a function of GMC masses
and discuss what mass pair is most likely to be observed.

This article is organized as follows.
In section~\ref{sec:previous_formulation}, we briefly review the
time-evolution equation on GMCMF formulated in \cite{Kobayashi2017a}.
In section~\ref{sec:sfccc_formulation},
we introduce our new time-evolution equation revised from \cite{Kobayashi2017a}
and introduce equations that calculate SFR\@.
In section~\ref{sec:results}, we explore our 
results:
GMCMF with triggered star formation in CCC sites, CCC-driven SFR,
and CCC frequency as a function of GMC masses.
The possible improvements
in our calculations are listed in section~\ref{sec:discussion}, 
and section~\ref{sec:summary} summarizes this article.
All the ``Log'' appeared in the figure labels are logarithm
in the base of 10.

\section{Basic Evolution Equation for Giant Molecular Cloud Mass Function}
\label{sec:previous_formulation}
Our formulation in this study is extensively based on our time evolution of GMCMF
from equation~(1) in \citet{Kobayashi2017a}.
Thus in this section, we briefly summarize \cite{Kobayashi2017a} and
refer the readers to \citet{Kobayashi2017a} for the detail assumptions
and formulations if necessary.

The time evolution of 
the differential number density of GMCs with mass $m$,
$n_{\rm cl}$, is evaluated as
\begin{eqnarray}
    \frac{\partial n_{\rm cl}}{\partial t} &+& \frac{\partial}{\partial m} 
    \left( n_{\rm cl} \left(\frac{{\rm d}m}{{\rm d}t}\right)_{\rm self} \right) \nonumber \\
    &=& -\frac{n_{\rm cl}}{\mytd} \nonumber \\
    &+& \frac{1}{2} \int_0^\infty \int_0^\infty  K(m_1, m_2) n_{{\rm cl},1} n_{{\rm cl},2} \nonumber \\
    & &~~~~~~~~~~~~~~~~~~~ \times \delta(m-m_1-m_2) {\rm d}m_1 {\rm d}m_2  \nonumber \\
    &-& \int_0^\infty K(m, m_2) n_{\rm cl} n_{{\rm cl},2} {\rm d}m_2
    +\left. \frac{1}{m} \frac{\partial \left(n_{\rm cl}m\right)}{\partial t} \right|_{\rm res} \,.
    \label{eq:coageq}
\end{eqnarray}
Here, $({\rm d}m/{\rm d}t)_{\rm self}$ represents the mass-growth rate of GMCs due to accretion from the ambient ISM,
$\mytd$ is the self-dispersal timescale of GMCs,
$n_{{\rm cl},1}$ and $n_{{\rm cl},2}$ are the differential number density
of GMCs whose masses are $m_1$ and $m_2$ respectively, 
$K(m_1, m_2)$ is the kernel function that 
determines the CCC rate between GMCs with mass $m_1$ and $m_2$,
$\delta$ is the Dirac delta function,
and $(1/m)\, \left. \partial \left(n_{\rm cl}m\right) / 
\partial t \right|_{\rm res}$
is the gas resurrection rate from dispersed gas.
Throughout this article, we assume that GMCs are molecular agglomeration 
bright in $^{12}$CO(1-0) line to be compared with observations.

We opt to employ $100$ pc as the disk scale height where GMCs populate,
which is observed in the Milky Way galaxy \citep[\eg,][]{Dame1987}.
The scale height observationally indicated has
a variation by a factor two to three (\eg, 35 pc \citep{stark2005}, 
half-luminosity height $\lesssim$ 60 pc \citep{Bronfman2000}),
thus CCC rate may increase by at most a factor two to three 
because a smaller thickness of the galactic disk 
means a larger number density of molecular clouds in the disk.  

\subsection{Self-Growth Term}
\label{subsec:growth}
The second term on the left-hand side of equation~(\ref{eq:coageq})
corresponds to a flux term in the conservation law.
The continuity equation in fluid dynamics
is one of such conservation laws, where
mass is the conserved quantity in the configuration space.
On the other hand, we here consider GMC number conservation in GMC mass space
because the number should be the conserved quantity
unless GMCs experience an abrupt change 
(\eg, dispersal or CCCs).
This term, therefore, 
represents
the GMC number flux in GMC mass space, 
which corresponds to GMC mass-growth in the configuration space.
The mass-growth rate $({\rm d}m/{\rm d}t)_{\rm self}$ can be basically evaluated
\begin{equation}
     \left(\frac{{\rm d}m}{{\rm d}t}\right)_{\rm self} = \frac{m}{\mytf} \,,
     \label{eq:selftf}
\end{equation}
where $\mytf$ is the typical mass-growth timescale.
This substitution is based on our assumption that 
the mass-growth rate (\ie, mass-gain rate from the ambient ISM)
is proportional to GMC's surface area and the surface area is
proportional to mass, if we employ the observational results that the majority of GMCs
have a similar column density $2 \times 10^{22} \, \mathrm{cm^{-2}}$ 
\citep{Onishi1999,Tachihara2000}.
The typical mass-growth timescale, $\mytf$, can have the same order of magnitude 
with the typical phase transition timescale if the mass-growth of molecular clouds
are driven by the phase transition from the ambient WNM to molecular gas
on the surface of GMCs, similar to the molecular cloud formation.
Such phase transition timescale is evaluated about a few 10 Myr,
over which molecular clouds successfully form out of magnetized WNM
by repeated
supersonic shocks from random direction due to 
expanding bubbles
(see \citet{Inutsuka2015,Kobayashi2017a}).
Therefore, we opt to choose $\mytf =10$ Myr as our fiducial timescale.
Our bubble paradigm is less likely to create large GMCs whose mass is 
comparable with or exceeds
the total gas mass that a single supernova
can sweep, thus we model such a cut-off mass scale $\sim 7 \times 10^6 \msun$
beyond which $\mytf$ becomes virtually 
infinite (see equation~(4) in \citet{Kobayashi2017a}).

Note that, in principle, $\mytf$ is the ensemble averaged timescale over different GMC 
mass-growth processes,
for which we here consider the multiple episodes of supersonic compressions is the most important
under the magnetic fields. The relative importance of different 
mass-growth processes depending on 
galactic environments needs to be further investigated in the future.

\subsection{Dispersal Term}
\label{subsec:dispersal}
The first term on the right-hand side of equation~(\ref{eq:coageq}) means the GMC self-dispersal due to
stellar feedback from massive stars born within those GMCs.
Here, this feedback can be 
any means (ionization, dissociation, heating, blowing-out, etc.).
The characteristic dispersal timescale, $\mytd$, can be evaluated as
\begin{equation}
    \mytd = \mytsb + \mytsd \,,
\end{equation}
where $\mytsb$ is the typical timescale for the protostars 
to evolve into the
main-sequence stars
after the birth of the concerned GMC, and $\mytsd$ is the typical timescale for
the complete destruction of GMCs after the star formation onset.
According to recent theories and observations within the Milky Way galaxy, 
the filamentary structure in densest parts of GMCs may host most of star formation
in GMCs
\citep[\eg,][]{Inutsuka2001,Andre2010,Andre2011,Roy2015},
and such filaments too can form through multiple supersonic shocks 
\citep[\cf,][]{Inoue2012,Inutsuka2015}. 
Therefore, we assume that $\mytsb$ would have a similar timescale as $\mytf$ so that 
we employ $\mytsb \sim 10$ Myr. According to line-radiation magnetohydrodynamics
simulations (\eg, results in \citet{Inutsuka2015} which updates \citet{Hosokawa2006a} 
by including magnetic fields), the typical timescale for the dissociation of CO molecules
is estimated as $\mytsd \sim 4$ Myr and is
irrespective of parental GMC mass
(see section~\ref{subsec:SFR} for the justification of this mass-independency argument).
Therefore, we expect that the typical dispersal timescale $\mytd = 10+4 = 14$ Myr.
Due to its definition, $\mytd$
essentially measures the typical
time-scale
over which 
GMCs are no longer identified in CO line observations. 
Thus this formulation implicitly allows 
the formation of CO-dark molecular gas (hereafter CO-dark gas)
whose population is left to be studied in the future. 

Note that $\mytd$ does not always guarantee complete blow-out of GMCs physically.
Meanwhile, several detailed semi-analytical studies 
\citep[\eg,][]{Kim2016a,Rahner2017} report that,
in some range of initial conditions of hydrogen number density in GMCs, 
GMC mass, and star cluster mass, 
stellar feedback from a single star cluster (both wind and radiation) cannot completely
blow out GMCs 
because of the gravity between 
the swept-up shell and the star cluster. This complete blow-out process
as well as CO-dark gas population also need to be investigated in the future.

Note that, similarly to $\mytf$, $\mytd$ is the ensemble averaged timescale 
over different GMC destructive processes.
Therefore, other processes may play an important role as well in different galactic environments; 
for example, galactic shear may dominate 
in much inner regions in galactic disks \citep[\cf][]{Dobbs2013}.
The relative importance of different destructive processes
needs to be further studied in the future (\cf, Jeffreson 2017 in prep.)

\subsection{Cloud-Cloud Collision Terms}
\label{subsec:ccc_term}
The second and third integration terms 
on the right-hand side of equation~(\ref{eq:coageq})
represent CCC, whose formulation is essentially the same as
the coagulation of two colliding dust particles in protoplanetary disks
\citep[\eg,][]{Trubnikov1971,Malyshkin2001}.
The first term in the two calculates the formation of GMCs with mass $m$
through the CCC between GMCs with mass $m_1$ and $m_2$.
The second term in the two calculates the formation of GMCs with mass $m+m_2$
through the CCC between GMCs with mass $m$ and $m_2$.
Therefore, to highlight the CCC effect simply,  
our formulation considers CCC as a coagulation process.
The CCC kernel function $K(m_1, m_2)$ is the product of
the total collisional cross section between GMCs with mass $m_1$ and $m_2$,
$\sigma_{{\rm col} \, 1,2}^{}$, and the relative velocity between the GMCs,
$V_{\rm rel}$,
\begin{equation}
    K(m_1,m_2) = \sigma_{{\rm col} \, 1,2}^{} \, V_{\rm rel} 
    = c_{\rm col}\frac{m_1 + m_2}{\Sigma_{\rm mol}^{}} V_{\rm rel,0} \,.
    \label{eq:kernel}
\end{equation}
Here, $c_{\rm col}$ is a correction factor, 
$\Sigma_{\rm mol}^{}$ is a typical column density of GMCs,
and $V_{\rm rel,0}$ is a typical relative velocity between GMCs.
Note that, in equations~(\ref{eq:coageq}) and~(\ref{eq:kernel}),
we restrict ourselves only to a perfect inelastic collision
case (\ie, coagulation) for simplicity.

The total collisional cross section can be essentially evaluated as
the total geometrical cross section of two colliding GMCs.
The GMC geometrical surface area can be estimated 
as their mass divided by a characteristic column density
$m/\Sigma_{\rm mol}^{}$,
given the observational fact that 
the majority of GMCs have 
a constant column density of
a few times $10^{22} \, \mathrm{cm^{-2}}$ if averaged over the entire cloud scale
(\eg, \citealt{Onishi1999,Tachihara2000}; see also subsection~\ref{subsec:corccc}
for its variation).
We opt to employ an observed value 
$\Sigma_{\rm mol}^{}= 2 \times 10^{22} \mu m_{\rm H} \mathrm{cm^{-2}}$,
where $\mu$ is the mean molecular weight and $m_{\rm H}$ is the 
atomic hydrogen weight.
Observationally, the cloud-to-cloud velocity dispersion 
is measured as
$8 - 10$ km s$^{-1}$ \citep{stark1989,stark2005,stark2006}.
The bubble paradigm predicts that 
GMCs are repeatedly
pushed by supersonic shocks
due to expanding shells and thus the sound speed of the medium 
within those expanding shells set the GMC velocity dispersion, 
which is about $10$ km s$^{-1}$.
Therefore, observed 
velocity dispersion
is consistent with our bubble paradigm
and we opt to set $V_{\rm rel,0} =10$ km s$^{-1}$.
Note that we turn off CCC calculations that involve GMCs whose 
cumulative number is less than 1, because such GMC populations
are less likely to exist in the real Universe (see also section 4 in
\cite{Kobayashi2017a} for the detail
and also subsection~\ref{subsec:corccc} of this article).

Several variations (\eg, gravitational focusing effect,
angle variation at which GMCs collide with each other) 
may make the total collisional cross section differ from
the total geometrical cross section. 
A factor of few differences due to these variations
may impact the GMCMF massive-end evolution 
and the computed total SFR
on the entire galactic disks,
but do not on the power-law slope
(see \citet{Kobayashi2017a}).
Thus, for simplicity, we opt to choose $c_{\rm col}=1$
(see also subsection~\ref{subsec:corccc} for other details involved in $c_{\rm col}$).

\subsection{Gas Resurrection}
\label{subsec:res}
The dispersal term in equation~(\ref{eq:coageq}) produces
dispersed gas. However, this term alone 
does not restore 
dispersed gas back into GMC populations. 
In reality when GMCs disperse, 
they turn into ambient ISM in several phases:
ionized, atomic, CO-dark, optically thick H{\sc i} etc.
Irrespective of phases, those dispersed gas may experience 
repeated 
supersonic shocks while floating around in the ISM 
to form a newer generation of GMCs
or to accrete onto pre-existing GMCs to help their mass-growth.
Hereafter, we call this process as ``gas resurrection''
following the nomenclature named in \citet{Kobayashi2017a}.
The last term in equation~(\ref{eq:coageq}) represents
this gas resurrection.
To calculate gas resurrection,
we introduce ``gas resurrecting factor'', $\myrec$, in \citet{Kobayashi2017a}, 
which is the mass fraction that is consumed to form newer generation of the minimum-mass 
GMCs out of the total amount of dispersed gas:
\begin{equation}
    \left. \frac{\partial \left(n_{\rm cl}m\right)}{\partial t} \right|_{\rm res} 
    = \myrec \dot{\rho}_{\rm total,disp} \delta(m-\mymmin) \,,
    \label{eq:fmin_res}
\end{equation}
Here, $\dot{\rho}_{\rm total,disp}$ is the total amount of dispersed gas produced 
from the system per unit time per unit volume, and $\mymmin$ is the minimum GMC mass (\ie, $10^4 \msun$
in this article). 
By this definition, in the steady state case, $1-\myrec$ fraction of the dispersed gas is
consumed to help the mass-growth of pre-existing intermediate-mass GMCs,
whose rate is given by the flux term $m/\mytf$ in equation~(\ref{eq:coageq}).

\subsection{Steady State Solution}
\label{subsec:steady}
\citet{Kobayashi2017a} reveal that the CCC impact is limited 
only in the massive-end evolution of GMCMF\@. Therefore, 
the power-law GMCMF feature in lower mass regime can be characterized by a 
steady state solution of the time-evolution equation
without the CCC terms as
\begin{equation}
    n_{\rm cl}(m) = n_0 \left( \frac{m}{\msun} \right)^{-1-\frac{\mytf}{\mytd}} \,.
    \label{eq:inutsuka_slope}
\end{equation}
Here, $n_0$ is the differential number density normalized at $m=\msun$.
This solution indicates that ongoing and future large radio surveys with higher spatial resolution
and higher sensitivity may constrain the timescale ratio $\mytf/\mytd$ 
by identifying smaller GMCs and measuring the power-law slope in GMCMF\footnote{In a crowded region such as galactic centers, 
the number density of GMCs is higher than in disk regions.
CCC may be a faster process than mass-growth or self-dispersal.
In such cases, the power-law slope varies with the dependence of 
the kernel function $K$ on GMC masses. See for this analysis in, for example,
equation~(A4) in \cite{Kwan1979} and equation~(31) in \cite{Kobayashi2017a}.}.

Equation~(\ref{eq:inutsuka_slope}) indicates that 
$\mytf$ can vary from 4 to 22 Myr to reproduce observed variation
in GMCMF slope given that $\mytd$ is presumably determined more by 
stellar evolution but not by galactic environment (\eg, arm or inter-arm;
see \citet{Kobayashi2017a}).
Indeed for example, based on PAWS data on Galaxy M51, \citet{Leroy2017} report that 
the depletion timescale due to star formation is almost constant with the total molecular column density 
in CO(1-0) line averaged on 40pc scale where the depletion timescale is 
defined as the total amount of molecular gas divided by SFR. Their derived depletion timescale 
$\sim 2$ Gyr and star formation efficiency $\sim 0.3$ per cent 
gives $\sim 6$ Myr as individual GMC dispersal timescale, 
which is a factor shorter than our fiducial dispersal timescale $\mytd = 14$ Myr.
This factor difference ($14/6=2.3$) 
may arise from shorter $\mytd$ in GMCs undergoing CCC
(see subsection~\ref{subsec:dispersal2}) but needs to be 
further investigated.

\section{REFORMULATION INCLUDING STAR FORMATION INDUCED BY CLOUD-CLOUD COLLISIONS}
\label{sec:sfccc_formulation}
In our previous time evolution equation introduced in equation~(\ref{eq:coageq}),
we do not implement any rapid star formation triggered by CCC\@.
However, observations of compact star cluster forming sites 
\citep[\eg][]{Torii2011,Kudryavtseva2012,Torii2015,Fukui2016,Fukui2017d,Kohno2017} 
indicate that GMCs are likely to form stars 
effectively (within a short timescale $\lesssim 1$ Myr) after GMCs experience CCC,
because of drastic compression of WNM and high accretion rate by enhanced sound velocity
\citep[\cf,][]{Inoue2017}.
Increasing number of CCC-candidate clouds reported from radio observations
\citep[\eg][]{Fukui2014,Fukui2016}
and the indication of frequent CCC events in 
galactic disk simulations
\citep[\eg,][]{Tasker2009, Dobbs2015}
suggest the importance in the investigation of 
the impact of CCC-driven star formation 
onto GMCMF evolution and 
its relative contribution to SFR for the entire galactic disks.

Before calculating the SFR, we first introduce a revised version of 
time evolution equation for GMCMF, by specifying the evolution 
of GMCs that are undergoing
the feedback from CCC-driven star cluster formation.
To do this, we subdivide GMC populations into two: the differential number density 
of GMCs of mass $m$ without experiencing CCC, $n_{\rm acc,cl}(m)$,
and the one with CCC experience, $n_{\rm col,cl}(m)$. 
Hereafter, we call the GMC populations in $n_{\rm acc,cl}(m)$ as ``normal'' GMCs
and the ones in $n_{\rm col,cl}(m)$ as ``CCC'' GMCs.
The total differential number density of GMCs with mass $m$, $n_{\rm cl}(m)$ is given as
\begin{equation}
    n_{\rm cl}(m) = n_{\rm acc,cl}(m) + n_{\rm col,cl}(m) \,.
\end{equation}
The basic evolution follows the same equation as equation~(\ref{eq:coageq}), but 
only CCC GMCs would have a shorter timescale for $\mytd$.
Thus the revised evolution equation becomes 
\begin{eqnarray}
    & &\frac{\partial \left(n_{\rm acc,cl}+n_{\rm col,cl}\right)}{\partial t} + 
    \frac{\partial}{\partial m} \left( (n_{\rm acc,cl}+n_{\rm col,cl}) \frac{m}{\mytf} \right) \nonumber \\
    &=& -\frac{n_{\rm acc,cl}}{\mytd} -\frac{n_{\rm col,cl}}{\mytdccc} \nonumber \\
    & &~ +\frac{1}{2} \int_0^\infty \! \int_0^\infty  K(m_1, m_2) \nonumber \\
    & &~~~~~~~~~ \times  (n_{{\rm acc,cl},1}+n_{{\rm col,cl},1})(n_{{\rm acc,cl},2}+n_{{\rm col,cl},2}) \nonumber \\
    & &~~~~~~~~~ \times \delta(m-m_1-m_2) {\rm d}m_1 {\rm d}m_2  \nonumber \\
    & &~ -\int_0^\infty K(m, m_2)  \nonumber \\ 
    & &~~~~~~~~~ \times (n_{\rm acc,cl}+n_{\rm col,cl}) (n_{{\rm acc,cl},2}+n_{{\rm col,cl},2}) {\rm d}m_2  \nonumber \\
    & &~ +\frac{1}{m} \left. \frac{\partial \left(n_{\rm cl}m\right)}{\partial t} \right|_{\rm res} \,. 
    \label{eq:coageq_withcccsf}
\end{eqnarray}
The subscripts $1$ and $2$ represent the mass bins $m_1$ and $m_2$ (\eg,
$n_{{\rm acc,cl},1}= n_{\rm acc,cl}(m_1)$).

This equation can be separated into two equations in which we calculate 
the time-evolution of $n_{\rm acc,cl}(m)$ and $n_{\rm col,cl}(m)$ respectively.
For normal GMCs,
\begin{eqnarray}
    & &\frac{\partial n_{\rm acc,cl}}{\partial t} + \frac{\partial}{\partial m} 
    \left(m \frac{n_{\rm acc,cl}}{\mytf}\right) \nonumber \\
    &=& -\frac{n_{\rm acc,cl}}{\mytd} \nonumber \\
    & &~ -\frac{n_{\rm acc,cl}}{n_{\rm cl}} \int_0^\infty K(m, m_2) \nonumber \\
    & &~~~~~~~~~ \times (n_{\rm acc,cl}+n_{\rm col,cl}) (n_{{\rm acc,cl},2}+n_{{\rm col,cl},2}) {\rm d}m_2 \nonumber \\
    & &~ +\frac{1}{m}\left. \frac{\partial \left(n_{\rm cl}m\right)}{\partial t} \right|_{\rm res} \,,
    \label{eq:coageq_normalGMC}
\end{eqnarray}
and for CCC GMCs,
\begin{eqnarray}
    & &\frac{\partial n_{\rm col,cl}}{\partial t} + \frac{\partial}{\partial m} 
    \left(m \frac{n_{\rm col,cl}}{\mytf}\right) \nonumber \\
    &=& -\frac{n_{\rm col,cl}}{\mytdccc} \nonumber \\
    & &~ +\frac{1}{2} \int_0^\infty \! \int_0^\infty  K(m_1, m_2)  \nonumber \\
    & &~~~~~~~~~ \times (n_{{\rm acc,cl},1}+n_{{\rm col,cl},1}) (n_{{\rm acc,cl},2}+n_{{\rm col,cl},2}) \nonumber \\
    & &~~~~~~~~~ \times \delta(m-m_1-m_2) {\rm d}m_1 {\rm d}m_2  \nonumber \\
    & &~ -\frac{n_{\rm col,cl}}{n_{\rm cl}} \int_0^\infty K(m, m_2)  \nonumber \\
    & &~~~~~~~~~ \times (n_{\rm acc,cl}+n_{\rm col,cl}) (n_{{\rm acc,cl},2}+n_{{\rm col,cl},2}) {\rm d}m_2  \,.
    \label{eq:coageq_CCCGMC}
\end{eqnarray}

We give the ample descriptions on each term in the following subsections.
All the parameters in this formulation are summarized in table~\ref{table:params}.

\begin{table}
    \caption{Parameters}
    \centering{
        \begin{tabular}{ccccc}
            \hline
            \hline
$\mytf$  &  $\mytd$  &  $\mytdccc$  &  $\myrec$  &  $\mysfe$ \\
$($ Myr $)$ &  $($ Myr $)$ & $($ Myr $)$ & & \\
 \hline 
 10               & 14  & 5       & 0.15          &  0.01 \\

            \hline 
            \hline
        \end{tabular}
    }\par
    \bigskip
    \textbf{Note.} Summary of the parameters in our formulation with their typical values.
              $\mytf$ denotes the mass-growth timescale of GMCs
              due to accretion of the ambient ISM by multiple episodes of compression.
              $\mytd$ is the GMC self-dispersal timescale due to the feedback by massive stars.
              $\mytdccc$ is similar to $\mytd$ but is determined by massive stars
              that CCC events form.
              $\myrec$ denotes the resurrecting factor
              (\ie the fractional mass out of total dispersed gas 
              that replenishes the minimum-mass GMC population).
              $\mysfe$ represents the fractional mass converted into stars in CCC sites from 
              their parental clouds.
              See also sections~\ref{sec:previous_formulation} and~\ref{sec:sfccc_formulation} 
              for the detailed descriptions.
    \label{table:params}
\end{table}

\begin{figure}\centering{\includegraphics[width=1.0\columnwidth]{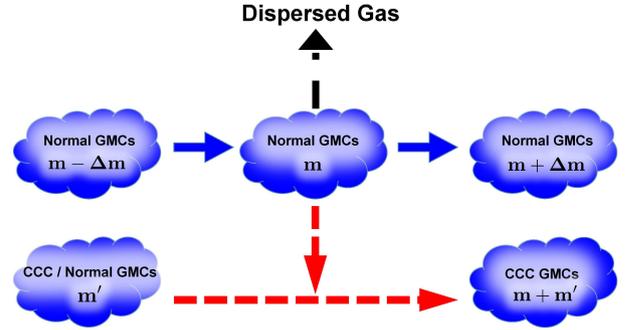}}
\caption{Flowchart 1 describing the mass-growth, self-dispersal, and CCC
of normal GMCs with mass $m$. 
The solid blue lines correspond to the mass-growth due to
multiple episodes of supersonic compressions. 
Given a mass bin width $\Delta m$ in calculation, the multiple compressions
grow GMCs from mass $m-\Delta m $ through $m$ to $m+\Delta m$.
The red dashed lines show the CCC process.
When normal GMCs with mass $m$ collide with GMCs with mass $m'$ (either in 
normal or CCC populations), they coagulate together to
create bigger GMCs with mass $m+m'$, which join CCC populations
but not normal populations.
The black dot-dashed lines are GMC self-dispersal.}
\label{fig:ccc_flow1}
\end{figure}
\begin{figure}\centering{\includegraphics[width=1.0\columnwidth]{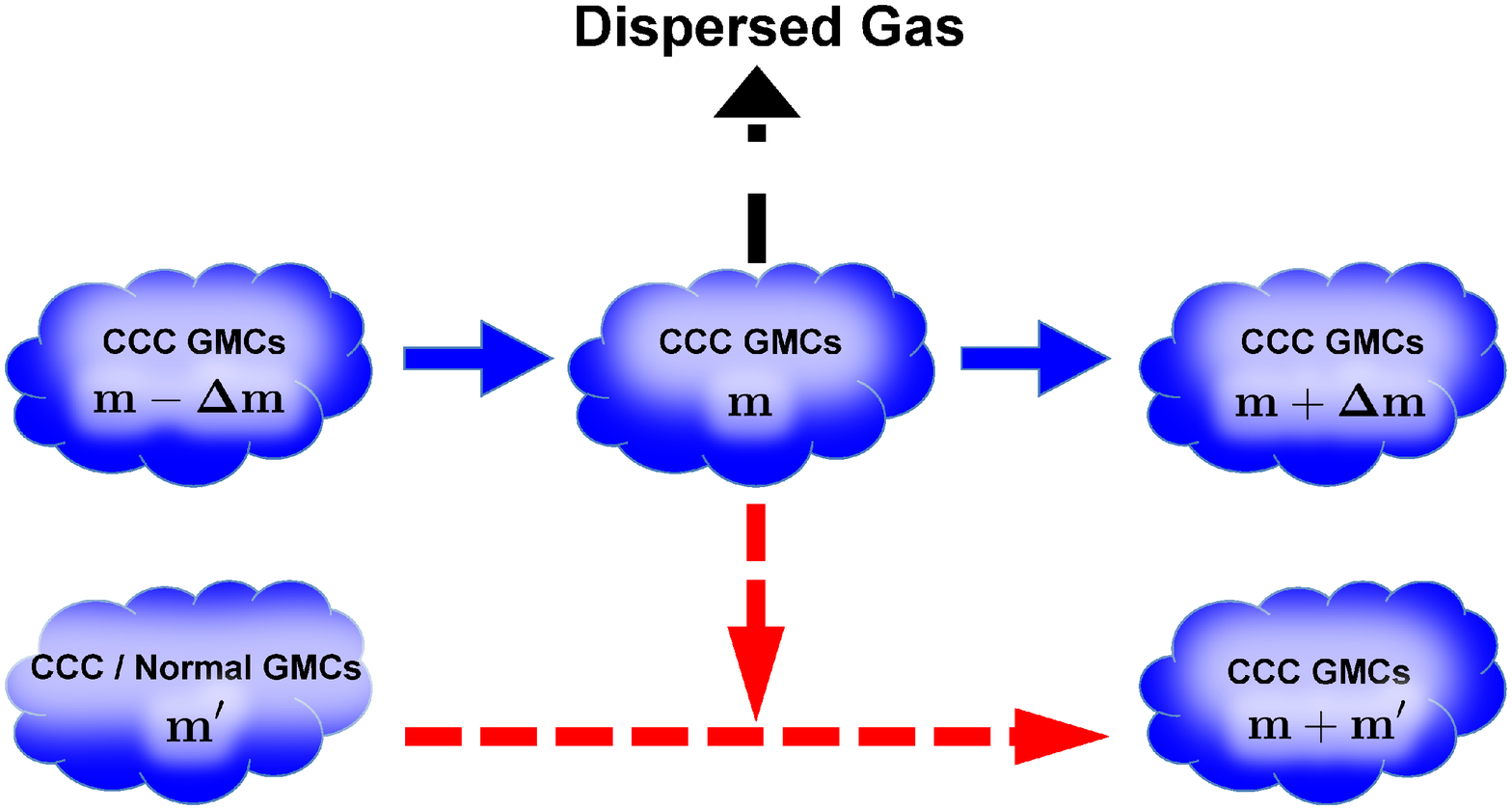}}
\caption{Flowchart 2 describing the mass-growth, self-dispersal, and CCC
of CCC GMCs with mass $m$.
The solid blue lines correspond to the mass-growth due to
multiple episodes of supersonic compressions. 
Given a mass bin width $\Delta m$ in calculation, the multiple compressions
grow GMCs from mass $m-\Delta m $ through $m$ to $m+\Delta m$.
Here, we assume that CCC GMCs remain in CCC populations through this mass-growth
because they are undergoing stellar feedback triggered by CCC,
which separates normal and CCC populations.
The red dashed lines show the CCC process.
When CCC GMCs with mass $m$ collide with GMCs with mass $m'$ (either in 
normal or CCC populations), they coagulate together to
create bigger GMCs with mass $m+m'$, which join CCC populations
but not normal populations.
The black dot-dashed lines are GMC self-dispersal, which have 
a shorter characteristic dispersal timescale compared with the one in normal populations
as discussed in subsection~\ref{subsec:dispersal2}.}
\label{fig:ccc_flow2}
\end{figure}

\subsection{Self-Growth Term}
\label{sub:growth2}
The second terms in equations~(\ref{eq:coageq_normalGMC}) and~(\ref{eq:coageq_CCCGMC}) 
correspond to GMC mass-growth due to multiple episodes of supersonic compression.
We assume that both normal population $n_{\rm acc,cl}(m)$ and CCC population $n_{\rm col,cl}(m)$ have the same 
$\mytf$ because the mass-growth driven by the phase transition dynamics 
presumably does not distinguish whether or not GMCs experience CCC\@.
Therefore, the mass-growth rate for both populations can be characterized as
$m/\mytf$ (see section~\ref{subsec:growth} for the justification
how $m/\mytf$ can be the mass-growth rate under the multiple episodes of supersonic compressions).
A schematic flowchart of this mass-growth is shown as blue solid lines in figures~\ref{fig:ccc_flow1}
and~\ref{fig:ccc_flow2}.

\subsection{Dispersal Term}
\label{subsec:dispersal2}
The first terms on the right hand side of equations~(\ref{eq:coageq_normalGMC}) and~(\ref{eq:coageq_CCCGMC})
represent GMC self-dispersal due to stellar feedback by massive stars born within GMCs.
Simulations of colliding GMCs \citep{Inoue2013,Takahira2014,Inoue2017}
suggest triggering
core formation in the shocked compressed layer.
Especially, \citet{Inoue2013} and \citet{Inoue2017} 
indicate that the
effective sound speed and resultant effective Jeans mass increase 
in the layer so that CCC enables rapid massive star formation.
In addition, observations suggest that GMCs undergoing CCC may form stars 
within a very short timescale $\lesssim 1$ Myr \citep[\cf][]{Kudryavtseva2012,Fukui2016}.
We therefore assume that, with a shorter star formation timescale $\mytsb = 1$ Myr, 
CCC GMCs have their dispersal timescale  $\mytdccc=\mytsb+\mytsd=5$ Myr.
A schematic flowchart of these dispersal processes is shown as black dot-dashed lines in figures~\ref{fig:ccc_flow1} 
and~\ref{fig:ccc_flow2}.

From the observational viewpoint, 
the stellar initial mass function (IMF) might be a top-heavy 
in cluster forming regions
(\eg, NGC6334: \citet{Munoz2007}, NGC3603:\citet{Harayama2008}).
Magnetohydrodynamics simulations also demonstrate
such top-heavy trend in CCC sites (at least 
before cores grow by mass accretion; \eg, \citet{Inoue2013}).
However for simplicity, we assume Salpeter IMF on the entire cloud scales even 
for GMCs undergoing or having undergone CCC\@. 
We opt to employ the same $\mytsd=4$ Myr for both normal and CCC GMCs
assuming that both GMC populations have the same dispersal efficiency with Salpeter IMF,
whereas the star formation timescale $\mytsb$ alone is shorter for CCC populations.

\subsection{Cloud-Cloud Collision Terms}
\label{subsec:ccc_term2}
The second term on the right hand side of equation~(\ref{eq:coageq_normalGMC})
and the last two terms in equation~(\ref{eq:coageq_CCCGMC})
correspond to CCC process. Equation~(\ref{eq:coageq_normalGMC}) has only one term 
because CCC process decreases but never increase the normal GMC populations.
Similarly to our previous study (\citet{Kobayashi2017a} and section~\ref{subsec:ccc_term}
in this article), we assume that CCC would work as a coagulation process
so that colliding GMCs essentially form a larger GMC\@.
Thus, the last term in equation~(\ref{eq:coageq_normalGMC}) represents the formation of 
CCC GMCs with mass $m+m_2$ through the CCC between GMCs with mass $m$ and $m_2$.
Similarly, the first CCC term in equation~(\ref{eq:coageq_CCCGMC})
represents the formation of CCC GMCs with mass $m$ through the CCC 
between GMCs with mass $m_1$ and $m_2$. 
Also, the second CCC term in equation~(\ref{eq:coageq_CCCGMC})
represents the formation of CCC GMCs with mass $m+m_2$ through
the CCC between GMCs with mass $m$ and $m_2$.
In this formulation, we assume a perfect inelastic collision for the CCC,
as we did in our previous formulation shown in Equation~(\ref{eq:coageq}).

We classify the resultant massive GMCs as CCC populations.
This treatment restricts ourselves to 
assuming that rapid star formation is always invoked
once GMCs experience CCC no matter what combination of GMC collide
(\ie, collisions between normal populations, CCC populations,
or normal and CCC populations).
In this manner, GMCs become quickly dispersed 
once they experience CCC with a shorter dispersal timescale $\mytdccc$ 
compared with normal GMCs.
A schematic flow of this CCC process is shown as red dashed lines 
in figures~\ref{fig:ccc_flow1} and~\ref{fig:ccc_flow2}.

Note that the CCC-driven star formation and subsequent stellar feedback
in our calculation does not create any smaller GMCs 
and thus CCC GMCs simply disperse at a given rate of $1/\mytdccc$.
The creation of such smaller GMCs by stellar feedback 
would impact the power-law slope in the low-mass regime,
which needs to be investigated further in the future.

\subsection{Gas Resurrection}
\label{subsec:res2}
The gas resurrection produces and replenishes the minimum-mass GMC populations. In this study, 
minimum-mass GMCs have only normal population but not CCC population because 
our CCC implementation does not produce any smaller mass clouds.
Therefore, the gas resurrection term appears only in equation~(\ref{eq:coageq_normalGMC})
but not in equation~(\ref{eq:coageq_CCCGMC}). 
This gas resurrection rate is calculated by equation~(\ref{eq:fmin_res}).
The mass production rate of dispersed gas, $\dot{\rho}_{\rm total,disp}$, 
should be computed from both normal and CCC GMC populations thus is computed as
\begin{equation}
    \dot{\rho}_{\rm total,disp} = \int \frac{m n_{\rm acc,cl}}{\mytd} {\rm d}m + \int \frac{m n_{\rm col,cl}}{\mytdccc} {\rm d}m \,.
\end{equation}
\citet{Kobayashi2017a} estimate that the steady state resurrecting factor for a typical galactic disk 
is about $\myrec = 0.15$ (\ie, 15 per cent gas resurrection).
We solve equation~(\ref{eq:coageq_withcccsf}) simultaneously with equation~(\ref{eq:fmin_res})
to calculate the gas resurrection.

\section{RESULTS}
\label{sec:results}
\subsection{Slope of Giant Molecular Cloud Mass Function}
\begin{figure}\centering{\includegraphics[width=1.0\columnwidth]{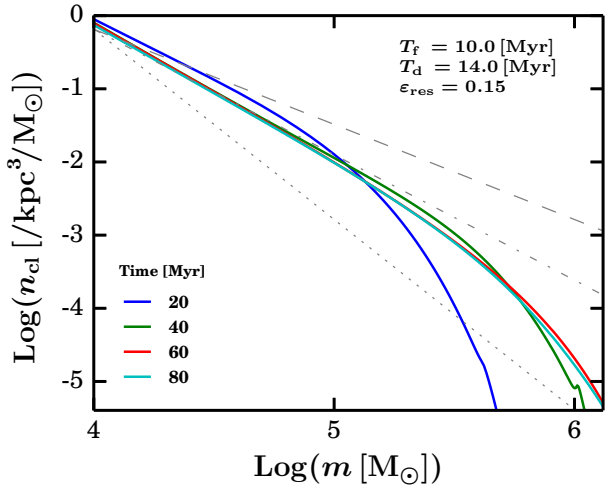}}
\caption{Differential number density $n_{\rm cl}$ as a function of
GMC mass, with $\mytf=10$ Myr, $\mytd=14$ Myr, $\mytdccc=5$ Myr, $\myrec=0.15$.
The color corresponds to time evolution.
As a reference, we plot three thin gray lines;
the dot-dashed line represents the steady state power-law slope 
$-1-\mytf/\mytd \sim -1.7$, the dashed line corresponds to the observed shallow slope
in arm regions of Galaxy M51,
and the dotted line corresponds to the observed steep slope in inter-arm regions of Galaxy M51.
The calculated GMCMF shows a power-law slope $\lesssim 10^{5.5} \msun$ close to the 
steady state slope indicated by equation~(\ref{eq:inutsuka_slope}).}
\label{fig:fid_gmcmf}
\end{figure}

We perform time integration of equation~(\ref{eq:coageq_withcccsf})
coupled with equation~(\ref{eq:fmin_res}).
Figure~\ref{fig:fid_gmcmf} shows the resultant time evolution of GMCMF.
We opt to employ our fiducial parameters 
(\ie, $\mytf=10$ Myr, $\mytd=14$ Myr, $\mytdccc=5$ Myr, $\myrec=0.15$).
The figure includes a reference dot-dashed line showing the steady state power-law slope
characterized by equation~(\ref{eq:inutsuka_slope}).
GMCMF in the mass range $m\lesssim 10^{5.5} \msun$ shows a single power-law slope
close to this steady state solution.
Compared with figure 7 in \cite{Kobayashi2017a}
where we calculated essentially the same condition but without CCC-driven star formation,
the number of massive GMCs $\gtrsim 10^6 \msun$ shown in figure~\ref{fig:fid_gmcmf} in this article
decreases due to star formation driven by CCC and subsequent stellar feedback.
Figure~\ref{fig:fid_gmcmf} also suggests that the power-law slope 
in the mass range $m\lesssim 10^{5,5} \msun$ is still preserved
over the GMCMF evolution even with CCC-driven star formation.
Therefore, our result indicates that CCC impacts only the massive-end of GMCMF.

\subsection{Star Formation Efficiency and Star Formation Rate}
\label{subsec:SFR}
To determine 
the relative contributions of normal and CCC GMCs onto
star formation on galactic scales, we need to calculate SFR in
each population. 
In this study, we opt to employ 
a given star formation efficiency (SFE) averaged over 
all GMC populations
to calculate SFR coarse-grained on galactic scale.

Hereafter, we define SFE, $\mysfe$, as the final mass fraction 
that goes into stars from a parental GMC
at the time when the entire GMC becomes completely dispersed.
Cumulative SFR can be evaluated as 
the product of SFE and 
the dispersal term in equation~(\ref{eq:coageq_withcccsf}):
\begin{eqnarray}
   \mathrm{SFR}(>m) &=& \mysfe \nonumber \\ 
   &\times& \left( \int_{m}^{\infty}\left. \frac{m n_{\rm acc,cl}}{\mytd}\right. \right. {\rm d}m 
   + \int_{m}^{\infty}\left. \left. \frac{m n_{\rm col,cl}}{\mytdccc}\right. {\rm d}m \right) \,. \nonumber \\
   & &
   \label{eq:csfr}
\end{eqnarray}
The first term corresponds to 
star formation due to normal GMC populations
whereas the second term represents 
star formation originated in CCC GMC populations.
For simplicity, we assume that the star formation timescale differs 
between two populations as included in $\mytd$ and $\mytdccc$ 
but the resultant SFE is the same for both populations as $\mysfe$.

Given a $n_{\rm cl}$ and $\mysfe$, one can calculate cumulative SFR with this equation.
On one hand, we evaluate $n_{\rm cl}$ directly from the calculated GMCMF\@.
On the other hand, we need to model $\mysfe$.
In principle, individual GMCs can have different SFE. 
Observationally, SFE averaged over a galactic disk
is equal to a few per cent
\citep{Zuckerman1974}.
In this study, we employ a fixed efficiency 1 per cent
as an ensemble-averaged SFE for simplicity.
This 1 per cent can be obtained as follows.
Given the Salpeter IMF, one massive star $\gtrsim 20 \msun$ can be born 
out of $1000 \msun$ star cluster. Such single massive star may disperse 
its parental cloud up to $10^5 \msun$ according to a detail line-radiation 
magnetohydrodynamics simulations \citep{Hosokawa2006a,Inutsuka2015}.
This suggests that SFE is 1 per cent on average 
(1000 $\msun$ star out of $10^5 \msun$ GMC).
This efficiency is essentially 
constant with
GMC mass because massive
GMCs $>10^{5} \msun$ create more massive stars and more dispersal.
Therefore, we employ $\mysfe=0.01$ as our fiducial value.

Note that, this mass-independent SFE is 
not applicable to the
low-mass GMCs $\lesssim 10^{5} \msun$ because 
their molecular gas mass is insufficient to produce a massive star
that can blow out the entire parental GMC
unless the stellar IMF in such low-mass GMCs prefer top-heavy IMF than Salpeter IMF\@. 
We are planning to investigate this effect and report in our 
forthcoming article.
Also note that we use $\mytd$ and $\mytdccc$ but neither $\mytsb$ nor
$\mytsd$ in equation~(\ref{eq:csfr}) because of our definition of SFE.

\begin{figure}\centering{\includegraphics[width=1.0\columnwidth]{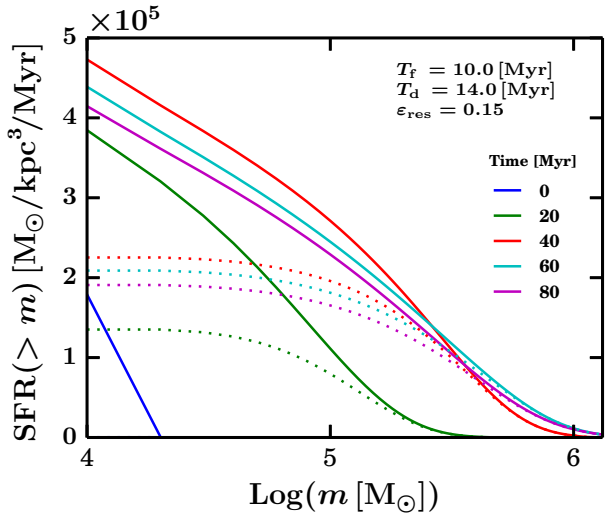}}
\caption{Cumulative star formation rate ${\rm SFR}(>m)$ as a function of GMC mass.
The color corresponds to time evolution. The solid lines represent the overall 
cumulative SFR originating from
normal and CCC GMCs, whereas the dotted lines
show the cumulative SFR originated only in CCC GMCs.
The vertical axis is in the unit of $\msun \,{\rm kpc^{-3}} \,{\rm Myr^{-1}}$,
thus total SFR on a galactic disk whose volume is similar to that of the Milky Way galaxy
(\eg, 10 kpc $\times$ 10 kpc $\times$ 100 pc)
is a few solar mass per year in the range plotted here.
This is a good agreement with observed typical SFR in the Milky Way
and nearby galaxies.}
\label{fig:fid_cSFR}
\end{figure}

Figure~\ref{fig:fid_cSFR} shows the time evolution of cumulative SFR
as a function of GMC mass in solid lines. 
This shows that the cumulative SFR becomes $\mathcal{O}(10^5) \msun \, \mathrm{kpc}^{-3} \, \mathrm{Myr}^{-1}$,
which corresponds to the typical SFR of a few solar mass per year over a galactic disk
(\eg, the Milky Way galaxy by Spitzer data \citet{Robitaille2010}).
In figure~\ref{fig:fid_cSFR},
we also plot the CCC-driven cumulative SFR in dotted lines,
which is a fraction of total cumulative SFR\@.
This suggests that most of the CCC-driven SFR 
comes from GMCs with mass $\gtsim 10^{5.5} \msun$, where the GMCMF
slope is significantly deviated from the steady state power-law slope.
In addition, the $\mathrm{SFR}(>10^4 \msun)$ indicates that 
the CCC-driven SFR may amount to a few 10 per cent (at most half) of 
the total SFR on galactic disk.
Our calculated CCC-driven SFR is presumably 
overestimated and may correspond to an upper limit 
because our formulation allows all colliding GMCs to coagulate together 
even when 
only their peripheries touch each other.
This overestimation is also due to 
assumed star formation efficiency in CCC GMCs
(see subsection~\ref{subsec:overestimated_SFR}).

As star formation goes, the GMC mass gradually accumulate into stars.
Our time-evolution equation (equation~(\ref{eq:coageq_withcccsf})) does not 
explicitly track such mass transformation. 
Although this is a very gradual process compared with other processes 
(mass-growth, dispersal, and CCC),
the such mass becomes $\sim 10^8 \msun$ accumulated over an entire galactic disk, 
if we integrate the evolution equation more than 100 Myr 
with a given SFR about a few solar mass per year. 
This may amount to at least a few per cent of the total molecular 
gas budget in a single galaxy. 
Therefore, to extend the current semi-analytical formulation to galaxy evolution over cosmological timescale,
mass transformation into stars needs to be formulated. 
We reserve this long-term evolution for future work.
In this case, we also must take into account the gas inflow from halos down to galactic disks,
which needs to be conducted together with cosmological large-scale structures.

\subsection{Cloud-Cloud Collision Frequency as a Function of GMC Mass}
\label{subsec:F_ccc}
\begin{figure*}
    \minipage[t]{0.5\textwidth}
        \includegraphics[scale=0.8]{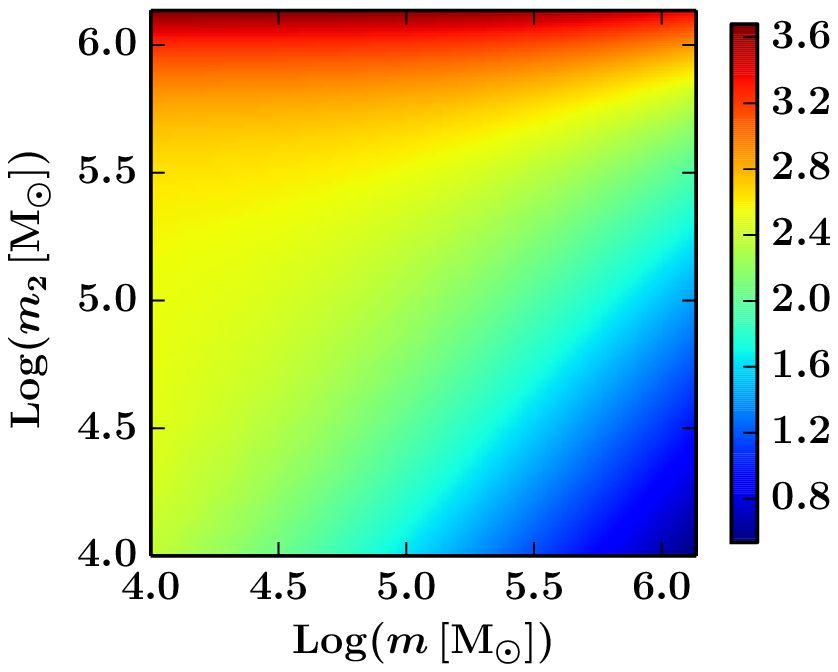}
    \endminipage%
    \minipage[t]{0.5\textwidth}
        \includegraphics[scale=0.8]{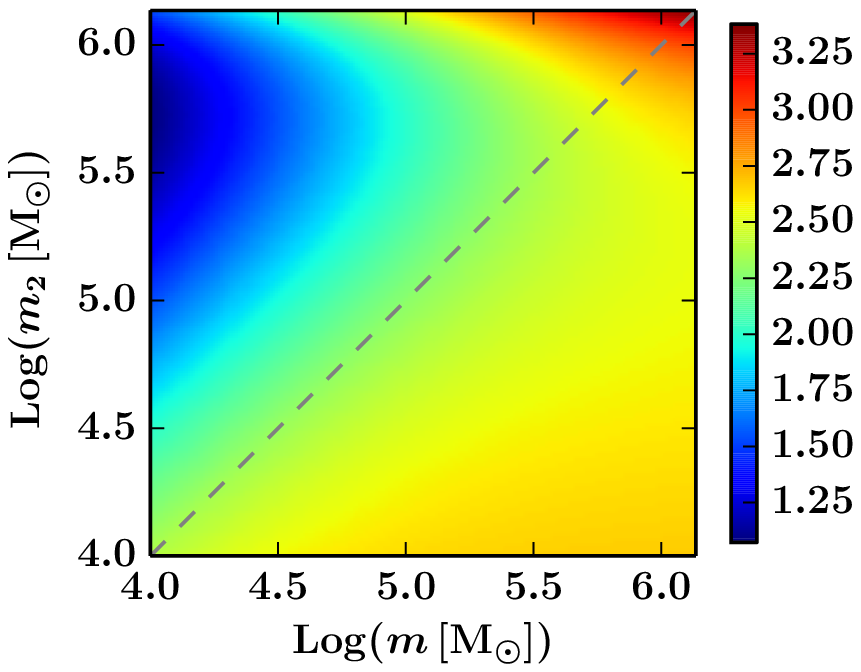}
    \endminipage%
    \caption{Left: The typical ``number collision timescale'' $\mytcolnum(m,m_2)$ 
    as a function of GMC mass combination involved in a CCC at $60$ Myr. 
    The horizontal axis corresponds to $m$ and the vertical axis corresponds to $m_2$.
    The color scale corresponds to $\log_{10}(\mytcolnum [\mathrm{Myr}])$. 
    Right: The typical ``mass collision timescale'' $\mytcolmass(m,m_2)$ 
    as a function of GMC mass combination involved in a CCC at $60$ Myr.
    The horizontal axis corresponds to $m$ and the vertical axis corresponds to $m_2$.
    The color scale corresponds to $\log_{10}(\mytcolmass [\mathrm{Myr}])$.
    The thin gray dashed line divides the panel into two regimes: $m>m_2$ (lower-right) and 
    $m<m_2$ (upper-left).
    }
    \label{fig:tcoli1i2}
\end{figure*}

\begin{figure}\centering{\includegraphics[width=1.0\columnwidth]{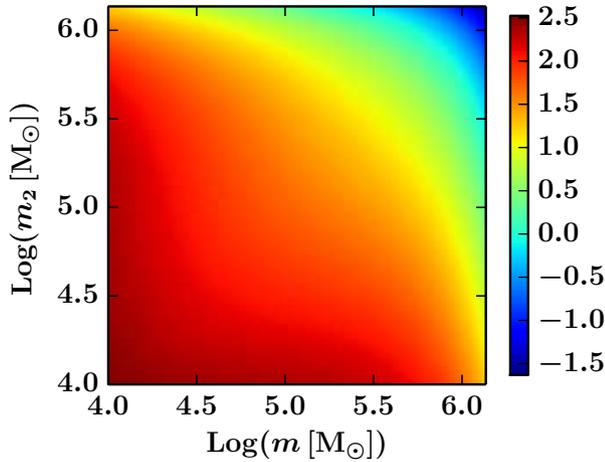}}
\caption{
    The expected number of CCC events observed in a galactic disk as a function of GMC mass pairs
    (\ie, $N_{\rm obs}$ defined by equation~(\ref{eq:n_event_ccc_obsability})).
    Here we opt to employ $\Delta T_{\rm obs} = 1 \mathrm{Myr}$
    and $V_{\rm survey} = 10 \mathrm{kpc^3}$ by assuming 
    an ideal condition where we can observe all the GMCs across
    the entire galactic disk.
    The color scale 
    is in the $\log_{10}$ scale.
    This figure indicates that 
    most of the observationally accessible CCC events 
    occur between GMCs with mass $\lesssim 10^{5.5}\msun$,
    whereas the CCC between massive GMCs, which presumably gives
    significant impact on CCC-driven star formation, may not be observed.
    }
\label{fig:tcoli1i2obs}
\end{figure}

In this subsection, we quantify the CCC frequency as a function of GMC mass.
We can define two different timescales that characterize CCC process:
``number collision timescale'' and ``mass collision timescale''.

Based on the collision term in equation~(\ref{eq:coageq_withcccsf}),
the total number of collisions that a single GMC with mass $m$ experiences
per unit time is given as the following integration:
\begin{eqnarray}
    &&\int  K(m, m_2) n_{{\rm cl},2} {\rm d}m_2 \nonumber \\
    &=& \int  K(m, m_2) n_{{\rm cl},2} m_2 {\rm d}\ln m_2 \,.
\end{eqnarray}
Therefore, 
the CCC event rate between a single GMC with mass $m$ and GMCs with $m_2$ with a given
differential number density 
$n_{{\rm cl},2}$ per unit logarithmic mass
interval $\Delta\ln m_2$ is 
\begin{equation}
    K(m, m_2) n_{{\rm cl},2} m_2 \,.
    \label{eq:CCCeventrate}
\end{equation}
One can evaluate the typical collision timescale for a single GMC with mass $m$ 
colliding with a GMC with mass $m_2$ as
\begin{equation}
    \mytcolnum(m, m_2) = \frac{1}{K(m, m_2) n_{{\rm cl},2} m_2} \,.
    \label{eq:tcoli1i2num}
\end{equation}
Let us call this timescale $\mytcolnum$ as ``number collision timescale'' because 
this is an e-folding timescale for the number of GMCs with mass $m$.
Similarly, the total mass-gain (\ie, mass-growth)
of a single GMC with mass $m$ due to CCC
is given as the following integration:
\begin{eqnarray}
    && \int  K(m, m_2) n_{{\rm cl},2} m_2 {\rm d}m_2 \nonumber \\
    &=& \int  K(m, m_2) n_{{\rm cl},2} m_2 m_2 {\rm d}\ln m_2 \,.
\end{eqnarray}
Therefore, we can also define another typical timescale, over which
a GMC with mass $m$ grows in mass 
due to CCC with GMCs of mass $m_2$ per unit logarithmic mass
interval $\Delta\ln m_2$ as:
\begin{equation}
    \mytcolmass(m, m_2) = \frac{m}{K(m, m_2) n_{{\rm cl},2} m_2 m_2} \,.
    \label{eq:tcoli1i2mass}
\end{equation}
Let us name this timescale $\mytcolmass$ as ``mass collision timescale'' because
this is an e-folding timescale for the total mass of GMCs with mass $m$.
The CCC frequency for a given GMC population with mass $m$ is therefore
characterized as a function of $m_2$ by $\mytcolnum(m, m_2)$ and 
$\mytcolmass(m, m_2)$.

Note that our formulation treats the CCC between GMCs with mass $m$ and $m_2$ as a coagulation
resulting into a GMC with mass $m+m_2$.
Therefore, $\mytcolmass(m,m_2)$ in the regime of $m \gg m_2$ represents the 
mass-growth timescale of GMCs with mass $m$, which corresponds to
the time evolution of GMCMF around mass $m$. 
However in the regime of $m \ll m_2$, the collisional outcome with mass $m+m_2$ 
is significantly larger than a GMC with mass $m$, thus 
$\mytcolmass(m,m_2)$ does not necessarily represent the time evolution of GMCMFs.
For example, a CCC between GMCs with mass $10^4 \msun$ and $10^6 \msun$ 
forms a GMC with mass $1.01 \times 10^6 \msun$. 
For the GMC with mass $10^4 \msun$,
this coagulation effectively looks like rapid mass-growth.
However, 
in terms of the GMCMF time evolution, this appears as the gradual mass-growth 
of GMCs at mass $10^6 \msun$.
It is thus more useful to compare $\mytcolmass(m, m_2)$ only in the regime of $m \geq m_2$
with other timescales (\eg, $\mytf$ and $\mytd$) 
when we discuss the time evolution of GMCMFs.

\subsubsection{Number Collision Timescale}
\label{subsubsec:Tcolnum}
The left panel in figure~\ref{fig:tcoli1i2} shows $\mytcolnum$
as a function of mass pair in one CCC event. 
Based on its definition in equation~(\ref{eq:tcoli1i2num}), $\mytcolnum (m,m_2)$
represents the e-folding timescale for a single GMC with mass $m$ due to the collisions
with GMCs with mass $m_2$. 
Thus the physical meaning of $\mytcolnum(m,m_2)$ is different from 
that of $\mytcolnum(m_2,m)$.
Indeed, figure~\ref{fig:tcoli1i2} shows such asymmetry between $m$ and $m_2$. 
Note that, $\mytcolnum(m,m_2)$ and $\mytcolnum(m_2,m)$ differ from
the total collisional event rate between $m$ and $m_2$,
which is symmetric between $m$ and $m_2$.
We discuss the total collisional event rate in section~\ref{subsubsec:observability}.

The figure indicates that massive GMCs (\ie, larger mass range in the horizontal axis) 
have higher opportunity to collide with smaller clouds
than with massive clouds because the number density of smaller clouds 
is larger than
that 
of massive clouds. Due to the same reason, smaller GMCs (\ie, smaller mass range in the horizontal axis)
also have higher opportunity
to collide with smaller clouds than with massive clouds.
Such intuitive understanding can be analytically confirmed by
equation~(\ref{eq:tcoli1i2num}) as follows. 
In $m\gg m_2$ regime, $\mytcolnum$ can be written as:
\begin{equation}
    \mytcolnum(m,m_2) \propto \frac{1}{m m_2^{1-\alpha}} \,.
    \label{eq:mytcolnum_trend1}
\end{equation}
Therefore $\mytcolnum$ becomes longer with $m_2$ 
as 
$\mytcolnum \propto m_2^{0.7}$
given the typical GMCMF slope $-\alpha \sim -1.7$
as shown in figure~\ref{fig:fid_gmcmf}.
This corresponds to the increasing trend of $\mytcolnum$
along the vertical axis at a given large mass in the horizontal mass coordinate 
in figure~\ref{fig:tcoli1i2}.
Note that $\mytcolnum(m,m_2)$ increases faster than $m_2^{0.7}$
in the range of $m_2>10^{5.5} \msun$. In this regime, 
the GMCMF deviates from the power-law distribution
assumed in equation~\ref{eq:mytcolnum_trend1}. Thus, 
$\mytcolnum(m,m_2)$ rapidly increases as $n_{\rm cl,2}$ decreases with $m_2$.
The resultant difference in $\mytcolnum$ is two or more orders of magnitude
(\eg, between $\mytcolnum(10^6 \msun, 10^4 \msun)$ and $\mytcolnum(10^6 \msun, 10^6 \msun)$).
On the other hand, in $m\lesssim m_2$ regime, $\mytcolnum$ can be evaluated as:
\begin{equation}
    \mytcolnum(m,m_2) \propto \frac{1}{m_2^{2-\alpha}} \,.
    \label{eq:mytcolnum_trend2}
\end{equation}
Therefore $\mytcolnum$ becomes shorter with $m_2$ as $\mytcolnum \propto m_2^{-0.3}$
given the typical GMCMF slope $-\alpha \sim -1.7$
as shown in figure~\ref{fig:fid_gmcmf}.
This corresponds to the decreasing trend of $\mytcolnum$
along the vertical axis at a given small mass in the horizontal mass coordinate
in figure~\ref{fig:tcoli1i2}.
However, its dependence on $m_2$ is limited to the power of $2-\alpha=0.3$
thus this trend is difficult to recognize in figure~\ref{fig:fid_gmcmf}.
In addition, similar to equation~(\ref{eq:mytcolnum_trend1}),
the power-law GMCMF assumption in equation~(\ref{eq:mytcolnum_trend2}) is 
invalid in the range of $m_2>10^{5.5} \msun$. Thus 
$\mytcolnum(m,m_2)$ increases rapidly as $n_{\rm cl,2}$ decreases with $m_2$.

Note that the above discussion is based on GMCMF with 
$-\alpha \sim -1.7$ assuming that this represents 
overall averaged GMC population on a galactic disk.
However, in case of inter-arm regions with $-\alpha < -2$ for example, 
$\mytcolnum$ is always increasing function with $m_2$ 
because the number of large clouds 
is very few.

\subsubsection{Mass Collision Timescale}
\label{subsubsec:Tcolmass}
As seen in section~\ref{subsubsec:Tcolnum}, 
$\mytcolnum$ characterizes 
the frequency of individual CCC events that a single GMC with mass $m$ experiences.
However this does not always characterize impact on GMCMF evolution,
because, for example, collision with small clouds may not largely 
contribute to mass-growth of massive clouds, which is not appreciable 
in GMCMF evolution.
Therefore, we evaluate the mass-growth driven by CCC
by calculating $\mytcolmass$, which gives the typical e-folding time
in mass for a single GMC with mass $m$ by CCC.
The right panel in figure~\ref{fig:tcoli1i2} shows $\mytcolmass$ 
as a function of GMC masses in a given GMC pair.
This panel indicates that 
the mass-growth of massive clouds
is still
dominated by CCC with small to inter-mediate mass clouds. However, 
such CCC events
increase only limited amount of
mass so that $\mytcolmass$ has 
only up to one order of magnitude difference from 
that of CCCs between
massive GMCs. 

These trends can be analytically confirmed by
equation~(\ref{eq:tcoli1i2mass}) as follows, similar to the discussion
for $\mytcolnum$. 
In $m\gg m_2$ regime, $\mytcolmass$ can be written as:
\begin{equation}
    \mytcolmass(m,m_2) \propto \frac{1}{m_2^{2-\alpha}} \,.
    \label{eq:mytcolmass_trend1}
\end{equation}
Therefore $\mytcolmass$ becomes shorter with $m_2$ 
as $\mytcolmass \propto m_2^{-0.3}$ given the typical GMCMF slope $-\alpha \sim -1.7$
as shown in figure~\ref{fig:fid_gmcmf}.
This corresponds to the decreasing trend of $\mytcolmass$
along the vertical axis at a given large mass in the horizontal mass coordinate 
in figure~\ref{fig:tcoli1i2}. Again, the rapid increment in $\mytcolmass$
in the range of $m_2>10^{5.5} \msun$ corresponds to the deviation of the GMCMF 
from the power-law distribution in this mass range.

Similarly, in $m\lesssim m_2$ regime, $\mytcolmass$ can be evaluated as: 
\begin{equation}
    \mytcolmass(m,m_2) \propto \frac{m}{m_2^{3-\alpha}} \,.
    \label{eq:mytcolmass_trend2}
\end{equation}
Therefore $\mytcolmass$ becomes shorter with $m_2$ 
as $\mytcolmass \propto m_2^{-1.3}$ given the typical GMCMF slope $-\alpha \sim -1.7$
as shown in figure~\ref{fig:fid_gmcmf}.
This corresponds to the decreasing trend of $\mytcolmass$
along the vertical axis at a given small mass in the horizontal mass coordinate
in figure~\ref{fig:tcoli1i2}.
The slight increment in $\mytcolmass$
in the range of $m_2>10^{5.5} \msun$ corresponds to the deviation of the GMCMF 
from the power-law distribution in this mass range.

Note that, as we have already discussed 
in the paragraph following equation~(\ref{eq:tcoli1i2mass}),
only the right-lower half of this panel (\ie, $m\geq m_2$ regime) 
can be directly compared with other timescales governing 
the GMCMF evolution (\eg, $\mytf$ and $\mytd$).
In this regime, 
the typical $\mytcolmass$ has the order of 100 Myr,
which is still longer than $\mytd$ or $\mytdccc$.
Therefore, the massive-end of GMCMF does not show significant growth after $60$ Myr.

\subsubsection{Observability}
\label{subsubsec:observability}
In this sub-subsection,
we explore simple estimation of CCC observability in galactic disks.
The observability of CCC between GMCs with mass $m$ and $m_2$,
$f_{\rm obs}(m,m_2)$,
can be characterized by multiplying its frequency and 
number density of GMCs:
\begin{equation}
    f_{\rm obs}(m,m_2) = \frac{\Delta n(m)}{\mytcolnum(m,m_2)} \,.
\end{equation}
The number density $\Delta n(m)$ 
can be estimated
from $n_{\rm cl}$ calculated in our GMCMF time evolution
as:
\begin{eqnarray}
    \Delta n(m) &=& \int_{m}^{m+\Delta m} n_{\rm cl} \, {\rm d}m \nonumber \\
                &=& \int_{\ln m}^{\ln(m+\Delta m)} n_{\rm cl} m \, {\rm d} \ln m \,.
\end{eqnarray}
This integration width $\Delta m$ is in principle determined by 
the capability
of individual observations.
Instead of specifying any capability,
we here employ the number density per unit logarithmic mass interval
$\Delta \ln m = \ln(m+\Delta m) - \ln m = 1$ for simplicity, thus
\begin{equation}
    \Delta n(m) = n_{\rm cl} m \,.
\end{equation}
This choice of $\Delta n(m)$ is consistent with the fact that we employ 
unit logarithmic mass interval $\Delta \ln m$ in equation~(\ref{eq:tcoli1i2mass}) 
to define $\mytcolnum$. This definition makes the observability $f_{\rm obs}(m,m_2)$ 
symmetric between $m$ and $m_2$ as:
\begin{equation}
    f_{\rm obs}(m,m_2) = K(m, m_2) n_{{\rm cl},2} m_2 n_{\rm cl} m \,.
\end{equation}
Such symmetry must exist because, unlike $\mytcolnum$, 
the number of total collisional events itself cannot distinguish between $m$ and $m_2$.

One can estimate the number of events expected to be observed within 
a galactic disk by multiplying $f_{\rm obs}$, the duration over which observation can 
identify colliding GMCs as a CCC event, $\Delta T_{\rm CCC}$, and the survey volume
$V_{\rm survey}$:
\begin{equation}
    N_{\rm obs} (m,m_2) = f_{\rm obs}(m,m_2) \Delta T_{\rm CCC} V_{\rm survey} \,.
    \label{eq:n_event_ccc_obsability}
\end{equation}
As a demonstration, we make a simple prediction for future surveys
under an ideal condition that we resolve and identify all the GMC populations
across an entire galactic disk.
Figure~\ref{fig:tcoli1i2obs} shows the resultant 
$N_{\rm obs}$ where we assume $\Delta T_{\rm CCC} \sim 1 \mathrm{Myr}$
and $V_{\rm survey} \sim 10 \mathrm{kpc^3}$, which corresponds to the 
total volume of Milky Way galactic thin disk in which GMCs most likely reside.
Our result suggests that we may observe over 100 events of CCC between $10^4 \msun$.
Indeed, most of the observed CCC candidates to date
involve $\mathcal{O}(10^4) \msun$ GMCs
\citep[\eg,][]{Fukui2016}.
On the other hand, 
our result also indicates that it 
is less likely to observe the CCC events between GMCs with mass $>10^{5.5} \msun$
that play a dominant role in CCC-driven SFR.

\section{DISCUSSION}
\label{sec:discussion}

\subsection{Correction Factor in Cloud-Cloud Collisions}
\label{subsec:corccc}
As shown in equation~(\ref{eq:kernel}) in section~\ref{subsec:ccc_term},
there is a correction factor $c_{\rm col}$ to evaluate the CCC rate.
There, we already mentioned two effects that are considered 
in this correction factor: 
gravitational focusing effect and angle with which GMCs collide each other.
In this subsection, we briefly introduce and explore several other factors 
about which the present authors 
are frequently asked. In the calculations we show in
previous sections, we assume that these factors 
cancel out each other for simplicity, because each factor 
either increase or decrease the CCC rate by a factor few.

\textit{Column Density of GMCs}:
Latest observations suggest that GMC mean column density 
could have its peak at $(2 - 6) \times 10^{21} \, \mathrm{cm^{-2}}$
(\eg, Auriga-California: \citet{Harvey2013}, Cygnus X: \citet{Schneider2016}, etc.).
This corresponds to visual extinction a few, which is able to protect 
CO molecules that radio surveys observe.
Our fiducial column density is $\Sigma_{\rm mol}^{} = 10^{22} \, \mathrm{cm^{-2}}$,
a factor few denser than the density suggested by the latest observations.
Therefore, the assumed CCC rate may be biased lower by a factor few.

\textit{Number Density and Relative Velocity}:
Observationally, both the number density and cloud-to-cloud velocity dispersion of GMCs 
increase towards galactic centers (\eg, Central Molecular Zone in the case of the Milky Way galaxy: 
\citet{Morris1996}). Therefore, the CCC rate is also enhanced at galactic centers.
Investigation in further inner region of galactic disks needs to take into account such variation.
However in this study, we restrict ourselves to 
a disk region, for example the solar circle in the case of the Milky Way galaxy.
Also, super star cluster formation
sites show high relative velocities
10 - 20 km s$^{-1}$ \citep[\eg,][]{Furukawa2009,Ohama2010,Fukui2014,Fukui2016},
where super star cluster is defined as star clusters having 10 - 20 O stars.
Such possibility that massive star formation is enhanced as a function of
relative velocity between GMCs also has to be investigated,
but we focus on 
a simple 
question
how much star formation can be induced by CCC at a given star formation efficiency in this study.
Note that, at galactic centers, 
the magnetic field strength was reported to be very large (\eg, a few milli-Gauss), 
at least, locally. 
This stronger fields 
presumably modifies self-growth timescale $\mytf$ as well.
Note also that, in non-disk small galaxies or some specific volume in galactic disks,
molecular cloud formation and subsequent star cluster formation can be 
triggered by large-scale colliding H{\sc i} flow 
\citep[\eg,][: inflow from Small Magellanic Cloud onto Large Magellanic Cloud]{Fukui2017a},
which needs to be investigated.

\textit{Area for One-Zone}:
Our time-evolution equation is essentially one-zone and
calculates the differential number density of GMC populations $n_{\rm cl}$,
which therefore does not require any specified three-dimensional configuration 
for concerned volume in which calculated GMCMFs exist under CCC-absent cases.
Nevertheless, one can still estimate calculated volume
that is self-consistent within our modeling.
For example, GMCs can travel roughly $1 \, \mathrm{kpc}$ in 
100 Myr with a proper velocity of $10 \, \mathrm{km \, s^{-1}}$.
Therefore, our calculated GMCMF up to 100 Myr should correspond to 
ensemble population of GMCs in a cylinder with a surface area of 
$1 \, \mathrm{kpc^2}$ and with a depth of $100$ pc given that 
a galactic thin disk has a scale height $100$ pc in which GMCs populates
(see section~\ref{sec:previous_formulation}).
It is less likely to have GMC collisions beyond this cylinder.
However, to enable comparisons with observations,
we assume that the area is bigger than $1 \, \mathrm{kpc^2}$ 
by referring to the observed area covered by subdivided regions
(\eg, PAWS \citet{Colombo2014a}:  from $7.54 \, \mathrm{kpc^2}$ in arm regions
to $19.99 \, \mathrm{kpc^2}$ in inter-arm regions),
because GMCs are statistically able to collide each other even if they are apart 
more than 1 kpc in each subdivided area.
In our calculation, this overestimated area overestimate the 
cumulative number of massive-end GMCs so that 
we invoke collisions with GMCs whose cumulative number is less than 1.
In this article, we aim at demonstrating 
our modeling for a typical region in a galactic disk
so that we opt to employ a cylinder with $10 \mathrm{kpc^2}$ area and with $100$ pc depth.

\subsection{Overestimation in Triggered Star Formation}
\label{subsec:overestimated_SFR}
As described in subsection~\ref{subsec:SFR},
we assume that SFE in CCC sites is 1 per cent ($\mysfe=0.01$)
of the coagulated parental GMC mass.
However, SFE
could be simply limited by 1 per cent of the smaller GMC in a CCC pair.
In case of a GMC pair with large mass difference,
masses $10^4 \msun$ and $10^6 \msun$ for example,
the resultant star cluster mass can be 
$\sim 10^2 \msun = 1 \%$ of $10^4 \msun$,
whereas our calculation estimates this as 
$\sim 10^4 \msun = 1 \%$ of $10^6 \msun$.
Therefore, in a CCC pair with large mass difference,
our SFE becomes close to 100 per cent 
out of smaller GMCs.

Radio observations of CCC sites \citep[\eg,][]{Fukui2016} suggest that 
at least 1 per cent of the smaller GMC mass in a given GMC pair
turns into massive stars $>20 \msun$ and
they could form as many massive stars equally as low-mass stars.
Detailed magnetohydrodynamics simulations 
\citep[\cf][]{Inoue2017} indicates that
molecular cloud cores may have a flat IMF,
although with poor statistics due to small number of samples.
Only if stellar population follows Salpeter IMF
even in CCC sites, then
the number of simultaneously-formed
low-mass stars $<20\msun$ 
is larger than that of
massive stars 
thus the total stellar mass 
can be comparable to colliding GMC mass, 
\ie, up to 100 per cent SFE of colliding GMC mass.

Therefore, if actual CCC sites preferentially form massive stars，
our calculated SFR triggered by CCC could be overestimated 
by a factor from a few to ten.  
This overestimation impacts the total SFR shown in figure~\ref{fig:fid_cSFR}
because CCC takes place most frequently between massive GMCs and small GMCs 
(see figure~\ref{fig:tcoli1i2}).

Coupled with possible overestimation in the CCC rate due to 
the perfect-inelastic assumption (see subsection~\ref{subsec:ccc_term2}),
we interpret our calculated SFR driven by CCC
as an upper limit.

\subsection{Lifetime and Age of GMCs}
\label{subsec:tlife_gmc}
The lifetime and age of GMCs are another two important quantities
to understand star formation and star cluster formation 
along with galaxy evolution (\eg, multiplicity of stellar ages in
individual star clusters, migration history of our solar system,
and so on).
We here distinguish  ``lifetime'' and ``age''of GMCs;
we define lifetime as 
the duration time for a GMC to its complete dispersal
once star formation starts to take place,
whereas age as the overall duration from GMC formation to its complete dispersal.

On one hand, the typical lifetime in our calculation corresponds to $\mytd=14$ Myr. 
Some observations indicate such 
short lifetime (20--30 Myr) inferred from GMC classification in Large Magellanic Cloud \citep{Kawamura2009}
and an upper limit (30 Myr) estimated by GMC number counts in inter-arm regions 
in Galaxy M51 \citep{Meidt2015}.
On the other hand, we have a delta function like mass distribution as the initial condition of 
our calculation
where the minimum-mass GMCs alone exist. This enables us to highlight how fast GMCs can grow.
To grow in mass, GMCs have to survive stellar feedback whose rate is determined by $\mytd$
(\ie, more massive GMCs have older ages).
Therefore, the age is in general longer than the lifetime.
Figure~\ref{fig:fid_gmcmf} suggests that the GMC age is $\gtsim 40 (80)$ Myr,
which GMCs require to grow from the minimum-mass
$10^4 \msun$ to $\gtsim 10^6 (10^7) \msun$. 
Such longevity is indicated by observations within the Milky Way galaxy
(\eg, \citealt{Barnes2011,Barnes2016} and Barnes et al 2017, ApJ, submitted; \cf, \citealt{Kauffmann2013})
and in nearby galaxies \citep[\eg,][]{Koda2009}.

By this analysis, we presume that observed short lifetimes are likely  
``lifetime'' determined by GMC dispersal rate, whereas observed long lifetimes mostly correspond
to ``age'' rather than lifetime.

\subsection{Background}
\label{subsec:reservoir}
In our current calculation, we assume that 
the background reservoir is always
plenty enough to sustain the steady state of GMCMF\@.
However, it is known in the Milky Way and nearby galaxies that the gas distribution 
(especially atom-to-molecular ratio) varies with the galactocentric radii
\citep[\cf,][]{Nakanishi2016}.
Therefore, it is desired to investigate the background gas evolution
coupled with galactic environment, which we reserve for future works.

\subsection{CCC-driven Star Formation}
\label{subsec:arminterarm_cccsf}
In the present article, we calculate and show CCC-driven SFR with a set of typical galactic disk parameters.
Our next subject is to 
compare CCC-driven SFRs between arm and inter-arm regions.
Intuitively, high CCC-driven SFR is expected in arm regions because the mass budget 
in arm regions is dominated by massive GMCs. However, mass-growth by multiple episodes of compression 
is also fast in arm regions (\ie, short $\mytf$) to quickly create large amount of normal GMC populations.
Thus, it is not obvious whether or not the ``fraction'' of CCC-driven SFR out of total SFR is high in arm regions,
and vice versa for inter-arm regions.
Indeed, our pilot calculations indicate that CCC-driven SFR covers 
$30$--$50$ per cent of star formation in arm regions
and $20$--$40$ per cent in inter-arm regions.
Time evolution of GMCMF and subsequent CCC-driven SFR have to be investigated further along with
the migration of GMC groups between different regions (\eg, arm to inter-arm and back into arm).
This involves time evolution in parameters (especially $\mytf$, $\mytd$, $\myrec$)
and is left for future studies.

\section{SUMMARY}
\label{sec:summary}
We have performed integration of time evolution equation for giant molecular cloud 
(GMC) mass functions including cloud-cloud collisions and subsequent star formation 
due to the collisions. 
Our results indicate that the stellar feedback triggered by 
cloud-cloud collisions modify only the massive end of GMC mass functions.
Thus the mass functions exhibit power-law slopes in the low-mass regime ($\lesssim 10^{5.5}\msun$), 
which can be characterized as a ratio of the GMC formation timescale to their dispersal 
timescale according to the environment in galactic disks.
The star formation rate (SFR) calculated with a given GMC mass function 
indicates that a few 10 per cent (at most half) of the galactic star formation 
may be operated by cloud-cloud collisions (CCCs).
This analysis is based on assumptions where
1) GMC population that experience cloud-cloud collisions have a
shorter star formation timescale than 
normal GMC population without cloud-cloud collisions,
and 2) both populations 
have the same star formation efficiency 
of 1 per cent with the Salpeter initial mass function.
Cloud-cloud collisions
may play a more important role if they would result in 
higher star formation efficiency.
Lastly, we also indicate that CCC between smaller clouds ($\sim 10^4 \msun$)
are more probable to be observed due to its large number,
although most of the CCC-driven SFR is triggered by CCC 
between massive GMCs $\gtsim 10^{5.5} \msun$.

\section*{ACKNOWLEDGEMENT}
We are grateful to the anonymous referee and our editor Asao Habe
for carefully reading our manuscript
and providing thoughtful comments, 
which improved our manuscript in great details.
MINK (15J04974), HK (26287101, 17K05632, 17H01105, 17H01103), 
SI (23244027, 16H02160), and YF (15H05694)
are supported by Grants-in-Aid from the Ministry of Education, Culture,
Sports, Science, and Technology of Japan. 
MINK thank 
Kengo Tachihara, Hidetoshi Sano, Atsushi Nishimura, Dario Colombo, Veselina Kalinova, Jens Kauffmann, Thushara Pillai,
Karl Menten, Eva Schinnerer, Sharon Meidt, Henrik Beuther, Juan Soler, Jouni Kainulainen,
Dimitrios Gouliermis, Maria Jesus Jimenez, Takuma Kokusho, and Doris Arzoumanian
for educating us with observational backgrounds. MINK also appreciate 
Tsuyoshi Inoue, Hosokawa Takashi, Kazunari Iwasaki,
Kengo Tomida, Kentaro Nagamine, Jonathan Tan, Stefanie Walch, Daniel Seifried,
Diederik Kruijssen, Melanie Chevance, Sarah Jeffreson, Daniel Haydon,
Marta Reina-Campos, Ralf Klessen, Eric Pellegrini, Daniel Rahner,
and Bhaskar Agarwal for fruitful discussion,
and Keiichi Kodaira, Hiroshi Karoji, Philippe Andre, Patrick Hennebelle, 
and Enrique Vazquez-Semadeni for encouraging MINK throughout this project.


\bibliographystyle{apj}
\bibliography{galaxy}

\end{document}